\documentclass[preprint,showpacs,aps,prb]{revtex4}
\usepackage{graphicx}
\usepackage{bm}

\begin{document}
\title{Landau-Zener transitions in a linear chain.}
\author{V.L. Pokrovsky$^{1,2}$, N.A. Sinitsyn$^1$}
\address{$^1$Department of Physics, Texas A\&M University, College Station, \\
Texas 77843-4242,\\
$^2$Landau Institute of Theoretical Physics, Chernogolovka, Moscow region\\
142432, Russia}
\date{today}

\begin{abstract}
We present an exact asymptotic solution for electron transition
amplitudes in an infinite linear chain driven by an external
time-dependent electric field. This solution extends the
Landau-Zener theory for the case of an infinite number of states
in the discrete spectrum. In addition to the transition amplitudes
we calculate the effective diffusion constant.
\end{abstract}
 \maketitle



Landau-Zener (LZ) theory \cite{landau},\cite{zener} treats a
quantum system placed in a slowly varying external field. If such
a system was prepared in a state of its discrete spectrum, it
adiabatically follows this state until its time dependent energy
level crosses another one. Near the crossing point the
adiabaticity can be violated and the system can escape from the
state it occupied initially to another one. Landau and Zener found
the transition probability for two-level crossing. The crossing of
more than two levels at the same time is generally an unlikely
coincidence. However, in some systems such a multi-level crossing
may occur systematically, due to the high symmetry of the
underlying Hamiltonian. The transition matrix for special cases of
multi-level crossing was studied in Refs.
\cite{{zeeman1},{zeeman2},{bow},{demkov},{deminf1},{elser}}.
Presently only a few exact results for multi-level crossing are
known. One of them relates to a multiplet of atomic electronic
states with a total spin $S$ or total rotational moment $J$ larger
than 1/2 in a varying external magnetic field
\cite{{zeeman1},{zeeman2}}. The Zeeman splitting between $2S+1 $
or $2J+1$ levels regularly vanishes at nodes of the magnetic
field. Another exactly solvable model displaying multi-level
crossing is the so-called bow-tie model \cite{bow}, whose physical
interpretation is not obvious.

Since its creation in 1932, LZ theory has had numerous
applications. They include molecular pre-dissociation \cite{LL},
slow atomic and molecular collisions \cite{collisions}, and
electron transfer in biomolecules \cite{biomolecules}. Recently
Wernsdorfer {\it et al.} \cite{WS1},\cite{WS2} employed the LZ
theory to describe consistently the step-like shape of the
hysteresis loop in special molecules with large magnetic moments
called nano-magnets. Using the LZ probability formula these
authors were able to find the extremely small tunnel splitting of
the classic degenerate ground states and even to reveal
oscillations of this value in external magnetic field. This
beautiful experiment, together with its clever treatment is a new
triumph of quantum mechanics and, in particular LZ theory.

The problem considered in this article is closely related to
another application of LZ theory: electronic transfer in
donor-acceptor complexes \cite{vol}. In this process of biological
and chemical importance, an electron tunnels between initial and
final positions through a long chain of identical sites. There are
two limiting cases for such a process. In the first case there is
no coherence between two sequential tunnelling processes
connecting nearest neighbor sites. In this case the probability of
tunnelling through several sites is very small in comparison to
that for one-site tunnelling. This limiting case was studied
earlier\cite{vol}. We consider the opposite limiting case in which
the sequential tunnelling processes are highly coherent and
tunnelling through many sites becomes available.

If the coherence between LZ transitions is lost, the problem is
reduced to multiplication and addition of probabilities, each
described by a proper LZ expression. The price we must pay for
incorporating the coherence between different transitions is a
strong reduction of the class of quantum systems considered. The
number of crossing levels in such systems must be infinite. The
hopping amplitudes from a site to its neighbors must be all
identical. Physically it describes the quantum electron transfer
between donor and acceptor separated by a long polymer strand
(molecular bridge). The bridge can be considered as a linear array
of identical sites. Such one-dimensional atomic-scale wires were
intensely studied, both experimentally and theoretically
\cite{{wire},{wire1},{wire2}}. Our result can be also applied to
transitions among electron states in semiconductor superlattices
\cite{{sl1},{sl2}}.

We study the tunnelling of a particle in such systems driven by a
time-dependent homogeneous external field.  An important
assumption is that all molecular fragments in the chain are
identical. An electric field splits the energy levels at different
sites of the chain and suppresses the transitions, which occur
within a narrow intervals about times when the electric field
becomes zero. Since the tunnelling is a fast process, we disregard
the oscillatory relaxation originating from phonons and other
elementary excitations.

Let denote $|n>$ a state located at the $n$-th site of the chain.
We assume that these states form a complete orthonormal set
(Wannier basis). In terms of this set the electron Hamiltonian
reads:

\begin{eqnarray}
\hat{H}= \sum\limits_{n=1}^{N} (\gamma\mid n\rangle\langle n +
1\mid + c.c.) + F(t) n\mid n\rangle\langle n\mid;\label{h0}\\
F(t)=eE(t)a \nonumber
\end{eqnarray}
where  $E(t)$ is the electric field, $e$ is the electron charge,
$a$ is the distance between sites and $\gamma$ is the coupling
constant (hopping amplitude). A series of exact solutions for the
time-dependent Shr\"odinger equation with the Hamiltonian
(\ref{h0}) for $N=\infty$, known as drifting plane waves, was
found long ago \cite{{sl2},{bloch},{wannier},{dev},{hid}}. Below
we solve the same problem for fixed initial conditions, thus
resolving the multi-state LZ problem.

The states $|n>$ are conventionally called the diabatic states.
They are the eigenstates of the diagonal part of this Hamiltonian
$H_0=F(t) n\mid n\rangle\langle n\mid$. The eigenstates of the
total Hamiltonian (\ref{h0}) depending on $t$ or $F(t)$ as
parameters are called adiabatic states. Until $\mid
F(t)\mid\gg\gamma$, the diabatic levels are close to the adiabatic
ones and the transitions between levels are suppressed. This is
the adiabatic regime. The adiabaticity is violated in the vicinity
of the electric field nodes determined by the inequality $\mid
F(t)\mid\leq\gamma$, where all transitions proceed. By level
crossing we mean that the diabatic levels cross, the exact
eigenvalues of the Hamiltonian (\ref{h0})  never cross, in
accordance to the Wigner - Neumann theorem. It is convenient to
place the time origin $t=0$ directly at the node of $F(T)$. Since
only a narrow vicinity of the node is substantial for transitions,
the exact dependence of the field on time can be reasonably
approximated by linear one: $F(t)\approx \dot{F}(0)\cdot t$. At
zero electric field $E$ and free boundary conditions the
Hamiltonian (\ref{h0}) can be diagonalized analytically. Its
spectrum is:
\begin{equation}
\varepsilon _j  = 2g\cos (\pi j/N);\,\,\,\,j=1 \ldots N
\label{exact}
\end{equation}

For non-zero field we have found the adiabatic eigenvalues
numerically. The result is shown in Fig. 1 for a finite chain
with 15 sites. For comparison the diabatic levels are depicted in
the same figure.
\begin{figure}
\includegraphics{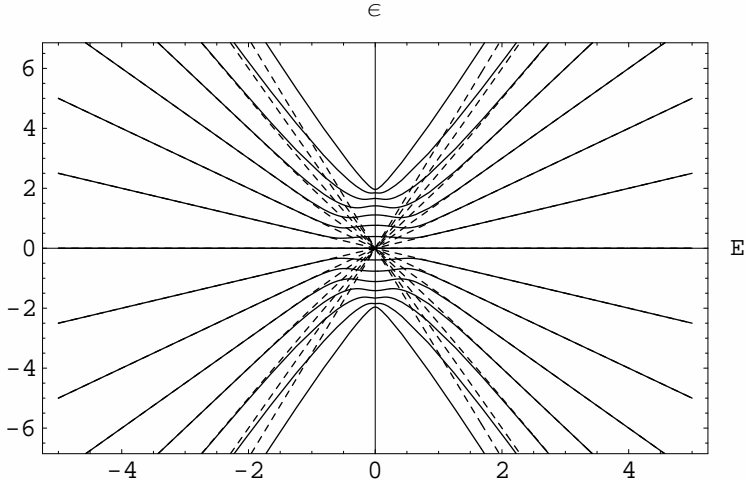}
\caption{Dependence of eigenvalues of the Hamiltonian (\ref{h0})
on $E$. Solid lines show adiabatic energies as function of $E$,
dashed lines depict diabatic energies. While the diabatic levels
intersect in one point, the adiabatic levels do not intersect.}
\label{Fig1}
\end{figure}
\noindent

We proceed to solve the time-dependent Schr\"odinger equation with
the Hamiltonian (\ref{h0}). Its matrix representation reads:
\begin{equation}
H_{nm}=n\dot{F}(0) t \delta_{nm}+\gamma
(\delta_{m,n+1}+\delta_{m,n-1}) \label{h3}
\end{equation}
For an infinite chain ($N\rightarrow \infty$) equation (\ref{h3})
is valid for all $n$ and $m$. After a proper rescaling of time the
Hamiltonian (\ref{h3}) becomes dimensionless:

\begin{equation}
H_{mn}=nt\delta _{mn}+g(\delta _{m,n+1}+\delta _{m,n-1})
\label{h4}
\end{equation}
It depends on only one dimensionless number, $g=\gamma
/\sqrt{\dot{F}(0)}$, which is the Landau-Zener parameter. Let the
time-dependent state vector be $\mid\alpha ,t\rangle=\sum_n
c_n(t)\mid n\rangle$. Then the system of equations for the
amplitudes $c_n(t)$ reads:

\begin{equation}
i\dot{c}_n=ntc_n+g(c_{n-1}+c_{n+1})  \label{z1}
\end{equation}
The transition matrix element $T_{n,n^{\prime}}$ should be
identified with the $t\rightarrow +\infty$ asymptote of an
amplitude $c_n(t)$ for a solution obeying the initial condition
$|c_m(t)|^2=\delta_{m,n^{\prime}}$ at $t\rightarrow -\infty$.
Since all $c_m(t)$ except of $c_n^{\prime}(t)$ are zero at
$t\rightarrow -\infty$, the initial condition can be more
explicitly written as
\begin{equation}
c_m(t\rightarrow -\infty)=\delta_{m,n^{\prime}}\exp{(-in^{\prime}t^2/2)}
\label{initial}
\end{equation}
We multiply the asymptotic values of $c_n(t)$ by $\exp{[int^2/2]}$
to remove strongly oscillating phase factors from
$T_{n,n^{\prime}}$.

Now introduce an auxiliary function $u(\varphi ,t)
=\sum_{n=-\infty }^{\infty}c_{n}(t)e^{in\varphi }$. The system
(\ref{z1}) is equivalent to the following equation in partial
derivatives for $u(\varphi ,t)$:
\begin{equation}
\frac{\partial u}{\partial t}+t
\frac{\partial u}{\partial \varphi }+2igu\cos
\varphi \,=\,0
\label{partial}
\end{equation}
The initial condition (\ref{initial}) is equivalent to the initial
condition: $u(\varphi ,t\rightarrow -\infty)\rightarrow \exp
[in^{\prime }(-\frac{t^{2}}{2}+\varphi )]$. Given the solution
$u(\varphi ,t)$, the amplitudes $c_{n}(t)$ can be found by the
inverse Fourier transformation $c_{n}(t)=\frac{1}{2\pi
}\int_{0}^{2\pi }u(\varphi ,t)e^{-in\varphi }d\varphi $. The
solution of eq. (\ref{partial}) that obeys proper boundary
conditions is:
\begin{eqnarray}
u(\varphi ,t)=\exp \left[ -i\left( 2g\int\limits_{-\infty }^{t}\cos \left(
\varphi -\frac{t^{2}}{2}+\frac{t^{^{\prime }2}}{2}\right) dt^{^{\prime
}}+\right.\right.\nonumber\\
\left.\left. n^{^{\prime }}\left( \varphi -\frac{t^{2}}{2}\right)
\right) \right]
\label{u}
\end{eqnarray}
Putting $t=+\infty$ in the solution (\ref{u}) and taking the
inverse Fourier-transform, we arrive at following asymptotic
values:
\begin{equation}
c_{n}(t)\approx \exp\left(-\frac{int^{2}}{2}+ i\frac{(n^{\prime
}-n)\pi}{4}\right)J_{|n-n^{\prime }|}(2\sqrt{2\pi }g) \label{ampl}
\end{equation}
Thus, the scattering amplitudes in terms of modified states, with
the fast phase factor $\exp (-int^{2}/2)$ incorporated, are:
\begin{equation}
T_{n,n^{\prime}}\equiv\langle n\mid T\mid n^{\prime }\rangle =
e^{i(n^{\prime }-n)\pi/4}J_{|n-n^{\prime }|}(2\sqrt{2\pi }g),
\label{scat}
\end{equation}
where the operator $T$ is expressed in terms of the evolution
operator $U(t,t^{\prime})$ for the Hamiltonian (\ref{h4}) in the
interaction representation:

\begin{equation}
T=\lim_{t\rightarrow\infty; \,  t^{\prime}\rightarrow -\infty}
\exp(i\int_{t^{\prime}}^0H_0(\tau)d\tau)U(t,t^{\prime})
\exp(i\int_0^tH_0(\tau)d\tau) \label{Uinft}
\end{equation}
The matrix elements $T_{n,n^{\prime}}$ display an infinite number
of oscillations with the LZ parameter $g$. However, for large
$|n-n^{\prime }|$ the oscillations start with $g>|n-n^{\prime }|$.
These oscillations can be observed experimentally by varying the
field sweep rate $\dot{E}(0)$. For small values of $g$ the
amplitudes are small and quickly decrease with growing
$|n-n^{\prime }|$. In Fig. 2 we depict transition probabilities
for several levels $n$ closest to the initial one $n^{\prime}$
versus the Landau-Zener parameter $g$. Fig. 3 shows the dependence
of the transition amplitude on $\mid n-n^{\prime}|$ at a fixed
value of $g$.

\begin{figure}
\includegraphics{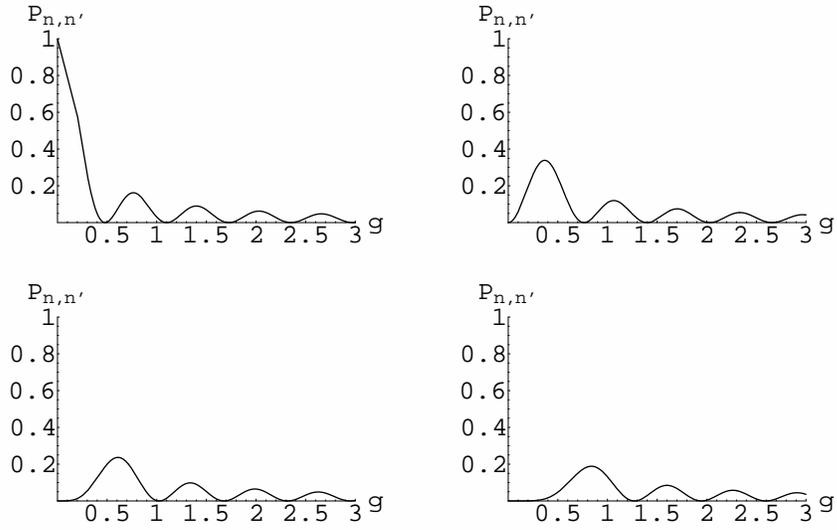}
\nopagebreak \caption{Transition probabilities to sites closest to
an initially filled one as functions of $g$: (a) $|n-n'|=0$; (b)
$|n-n'|=1$; (c) $|n-n'|=2$; (d) $|n-n'|=3.$} \label{Fig2}
\end{figure}
\noindent

\begin{figure}
\includegraphics{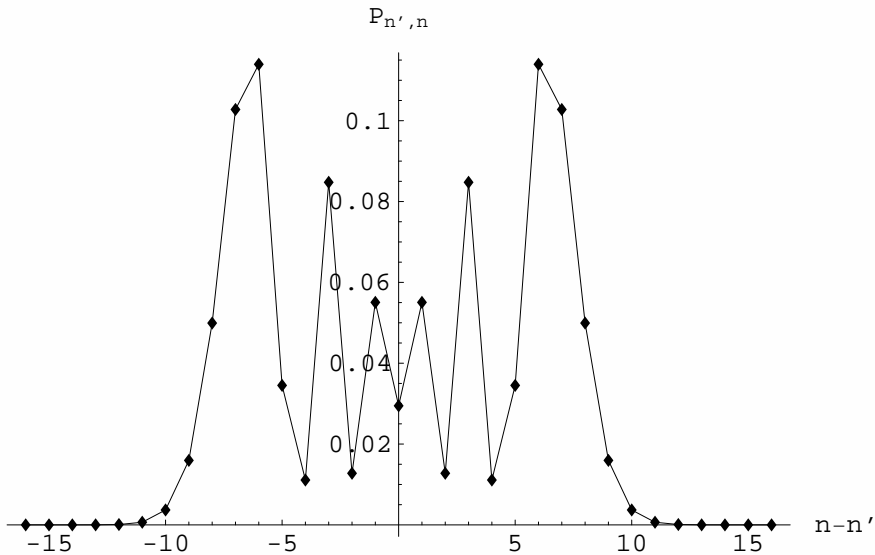}
\caption{Transition probabilities vs. $n-n'$ at a fixed $g=1.6.
\, Solid \, line \, is \, a \, guide \, for \, eye.$}
\label{Fig3}
\end{figure}
\noindent

For large $g \gg |n-n^{\prime}|$ the asymptotic values of the
amplitudes (\ref{scat}) are:

\begin{equation}
\langle n\mid T\mid n^{\prime }\rangle \sim \frac{e^{i(n^{\prime
}-n)\pi /4}}{(\sqrt{2\pi^3}g)^{1/2}}\cos(2\sqrt{2\pi
}g-\frac{(n-n')\pi}{2}-\frac{\pi}{4})
\label{lg}
\end{equation}

It is instructive to compare this result with other exactly
solvable generalized Landau-Zener models. Most of them refer to
systems with a finite number of states $N$. In the limit $g \gg N$
the transition probabilities behave like an exponent
$\exp(-C(n,n') g^2)$, where the $C(n,n')$ do not depend on $g$. In
contrast, the result (\ref{lg}) displays a power law with
oscillations instead of an exponential dependence on $g$ for large
$g$. This is the manifestation of quantum interference of
different Feynman trajectories, which are discrete in the chain. A
step in a trajectory has average length $g$ (see below). Such a
step cannot be realized in a system with a finite number of states
if $g\gg N$.

The mean square displacement at one crossing event is:

$$
<(n-n')^2>= \sum \limits_{n=-\infty}^{\infty} (n-n')^2
|J_{|n-n^{\prime }|}(2\sqrt{2\pi }g)|^2 =
$$
\begin{equation}
=4\pi g^2
\label{nmean}
\end{equation}
If the external field is periodic in time and the coherence
between crossing events is lost\footnote{Do not confuse the
decoherence between different crossings events with the
decoherence between tunnelling on different distances during one
crossing.}, the electron performs a random walk, i.e. it diffuses.
Assume the field to oscillate harmonically as $F(t)=F_0 \sin
(\omega t)$. At the nodes $t_k=\pi k /\omega$ ($k$ is an integer)
all diabatic levels cross together. The squared Landau-Zener
parameter is $ g^2=\frac{\gamma ^2}{ F_0\omega}$. The diffusion
coefficient is $ D=2a^2 <(n-n')^2>/T $, where $T=2\pi/\omega$ is
the period of oscillations and the factor $2$ accounts for two
crossing events per period. Collecting these results and equation
(\ref{nmean}), we find:
\begin{equation}
D=\frac{4 a \gamma ^2}{F_0} \label{dif1}
\end{equation}
This result does not depend on the frequency of the external field.

The theory can be extended to a more general Hamiltonian
incorporating hopping between any two sites, but conserving
translational invariance:

\begin{eqnarray}
H=\sum_{m,n}H_{m,n}|m\rangle\langle n|;\nonumber\\
 H_{mn}=nt\delta _{mn}+g_{|m-n|};\quad g_{-k}=g_{k}^{*}
\label{ext}
\end{eqnarray}
For simplicity we present below the result for real hopping
amplitudes $g_{k}=g_{-k}$:

\begin{eqnarray}
\left\langle n\right| T\left| n^{^{\prime }}\right\rangle =\frac{%
e^{i(n^{\prime }-n)\pi /4}}{2\pi }%
\int \limits_{0}^{2\pi }\exp \left( -i2\sqrt{2\pi }f(\varphi
,g_{j})\right.\nonumber\\
\left. +i(n^{^{\prime }}-n)\varphi \right) d\varphi
\label{amplext}
\end{eqnarray}
where $f(\varphi ,g_{j})=\sum\nolimits_{k}\frac{g_{k}}{\sqrt{k}}\cos
k\varphi $.

The model (\ref{h0}) can be generalized also to incorporate
internal degrees of freedom of identical chain fragments. In this
case the local states are described by amplitudes $a_{n,\alpha}$
with two indices. The first index $n$ denotes the position and the
second index $\alpha$ labels the inner states. The Schr\"odinger
equation for the amplitudes then reads:

\begin{equation}
i\dot{a}_{n,\alpha}=(nt+{\epsilon}_{\alpha}+{\delta}_{\alpha}t)a_{n,\alpha}+\sum
\limits_\beta g_{\alpha,\beta}(a_{n+1,\beta}+a_{n-1,\beta})
\label{more}
\end{equation}

\noindent where the indexes $\alpha,\beta=1 \ldots N_{int}$ run
over the internal states of the molecular wire segment. Changing
to variables $a_{n,\alpha}= b_{n,\alpha} e^{-int^2/2}$ we
eliminate the term proportional to $t$ in equation (\ref{more}).
Introducing a new function
$u_{\alpha}(\varphi)=\sum_{n=-\infty}^{\infty}b_{n,\alpha}e^{in\varphi}$,
we reduce the infinite system (\ref{more}) to a finite set of
$N_{int}$ ordinary differential equations:

\begin{equation}
i\dot{u}_{\alpha}=(\epsilon_\alpha+{\delta}_{\alpha}t)
u_{\alpha}+2\cos (t^2+\varphi)\sum \limits_\beta
g_{\alpha,\beta}u_\beta
\label{more2}
\end{equation}
in which $\varphi$ plays the role of a parameter. The initial
conditions are $b_{n,\alpha}(t\rightarrow -\infty)=\exp{(-i
\delta_{\alpha}t^2/2)} \delta_{n,n^{\prime}}$ and
$u_{\alpha}(\varphi,t\rightarrow -\infty
)=\exp{[(-i\delta_{\alpha}t^2/2)+in^{\prime}\varphi]}$. Thus, the
variable (parameter) $\varphi$ enters not only in the system
(\ref{more2}), but also in the initial conditions. This system
must be solved for all values of parameter $\varphi$ in the
interval $(0,2\pi)$. The inverse Fourier-transformation yields the
evolution operator just as for the case $N_{int}=1$. An analytical
solution of the system (\ref{more2}) is possible for some special
choices of the parameters $\varepsilon_{\alpha}, \delta_{\alpha}$
and $g_{\alpha,\beta}$. For example, two identical coupled chains
correspond to $N_{int}=2$. Then the indices $\alpha,\beta$ take on
the values 1,2. The simplest solvable choice of parameters is:
$\varepsilon_1=0, \varepsilon_2=\varepsilon; \delta_1=0,
\delta_2=\delta; g_{1,1}=g_{2,2}=\gamma;
g_{1,2}=g_{2,1}=\gamma^{\prime}$. Exact solution of this model can
be reduced to ones solved in this article together with solvable
two level LZ model.

In conclusion, we have generalized the LZ theory to an infinite
number of crossing levels. Physically it describes an electron on
an infinite chain subject to a time-dependent electric field. The
high symmetry of the problem allows us to find not only the
asymptotics, but also the intermediate values of the amplitudes.
Our solution is valid for an infinite chain with translational
symmetry. We demonstrated it for the case of a simple primitive
cell, but it can be generalized and in some cases exactly solved
even if the primitive cell contains more than one site.

Finite size effects do not permit us to apply our result
(\ref{scat}) directly to the transition amplitude from the first
to the last site of the chain even if it contains many sites. In
particular, the transition amplitude from one end to another does
not oscillate. However, our calculation of the diffusion
coefficient (\ref{dif1}) is valid since diffusion presumably
proceeds far from the ends of the chain. Certainly, we assume that
coherence is lost during the time interval between two sequential
crossing events.

Our solution demonstrates a phenomenon that is probably common for
most systems with multi-level crossing: oscillations of the
transition probabilities as a function of the LZ parameter and
site position (distance between diabatic levels). However, their
asymptotic values for large values of the LZ parameter differ from
those for other solvable multi-state LZ-models with a finite
number of states. We expect that in a general situation with $N
\gg 1$ crossing levels, the transition probabilities will behave
similarly to those found in this work for $1 \ll g \ll N$,
provided that the initially occupied states are far enough from
the diabatic spectrum boundaries.

Finally we discuss the relationship between our problem and a
typical problem for semiconductor superlattices
\cite{{sl1},{sl2}}. The latter is associated with Anderson
localization. The diabatic levels at sites are randomly
distributed. In one and two dimensions all sites are localized. If
the width of the energy distribution $\Delta$ is much less than
the tunnelling amplitude $\gamma$, the localization length in one
dimension is $a\gamma /\Delta$. To enhance the tunnelling through
a chain it is reasonable to apply a time-dependent electric field.
The electric field is substantial if $F\gamma /\Delta \geq \Delta$
where $F$ is a typical value of $F(t)$. Our approximation is valid
if the inequality is strong: $F\gamma /\Delta \gg \Delta$.
Tunnelling transitions in the field proceed during an interval of
time $\tau$ defined by relation $F(\tau)\sim\gamma$. The value
$F(\tau)$ can be accepted for $F$. We see that the strong
inequality $\gamma \gg\Delta$ guarantees the existence of the
strong field limit in which the randomness of levels can be
ignored. This requirement does not impose any limitations on the
LZ parameter $g$.

\section{acknowledgements}
We thank W.Saslow for critical reading of the manuscript. This
work was supported by NSF under the grant DMR 0072115 and by DOE
under the grant DE-FG03-96ER45598. One of us (VP) acknowledges the
support of the Humboldt Foundation.

\end{document}